\documentclass[a4paper,12pt]{article}
\usepackage{mathptmx}
\usepackage{epsfig}
\usepackage{float}
\usepackage{amssymb}
\usepackage{latexsym}
\usepackage{authblk}
\usepackage{natbib}
\usepackage{graphicx}
\usepackage{multirow}

\usepackage{amsfonts}
\newcommand{\beq}{\begin{equation}}
\newcommand{\eeq}{\end{equation}}
\newcommand{\beqa}{\begin{eqnarray}}
\newcommand{\eeqa}{\end{eqnarray}}

\usepackage{natbib}
\usepackage{graphicx}
\usepackage{multirow}
\begin{document}
%\begin{frontmatter}

%\maketitle
\title{Current status and possible extension of the global neutron monitor network}
\author[1,2]{A.L. Mishev}
\author[1,2]{I.G. Usoskin}
\affil[1]{Space Physics and Astronomy Research Unit, University of Oulu, Finland.}
\affil[2]{Sodankyl\"a Geophysical Observatory, University of Oulu, Finland.}
\maketitle
\begin{abstract}

The global neutron monitor network has been successfully used over several decades to study cosmic ray variations and fluxes of energetic solar particles. Nowadays, it is used also for space weather purposes, e.g. alerts and assessment of the exposure to radiation. Here, we present the current status of the global neutron monitor network. We discuss the ability of the global neutron monitor network to study solar energetic particles, specifically during large ground level enhancements. We demonstrate as an example, the derived solar proton characteristics during ground level enhancements GLE $\#$5 and the resulting effective dose over the globe at a typical commercial jet flight altitude of 40 kft ($\approx$ 12 200m) above sea level. We present a plan for improvement of space weather services and applications of the global neutron monitor network, specifically for studies related to solar energetic particles, namely an extension of the existing network with several new monitors. We discuss the ability of the optimized global neutron monitor network to study various populations of solar energetic particles and to provide reliable space weather services.  
\end{abstract}

\small Keywords:Solar energetic particles, GLE events, neutron monitor network, radiation environment 
 \normalsize

\label{cor}{\small For contact: alexander.mishev@oulu.fi;alex\_mishev@yahoo.com}

%\end{frontmatter}

\section{Introduction: science background and motivation}
Cosmic rays (CRs) represent flux of high-energy subatomic particles, mostly protons, $\alpha$-particles and traces of heavier nuclei. Their energy ranges from about $10^{6}$ to $10^{21}$ eV, following a power-law spectrum \citep{Tanabashi2018}. The bulk of CRs originate from the Galaxy, called galactic cosmic rays (GCRs), produced during and/or following supernova explosions, e.g., in supernova remnants. GCRs are always present in the vicinity of the Earth and permanently impinge on the Earth's atmosphere. While the low-energy CR particles are absorbed in the upper atmosphere, those with energies about GeV nucleon$^{-1}$ produce secondary particles via interactions with the atmospheric atoms. Those secondaries also collide with air nuclei, in turn producing other particles, if their energy is sufficiently high. Each collision adds a certain amount of particles, leading to the development of a complicated nuclear-electromagnetic-muon cascade known as an extensive air shower \citep[for details see][and references therein]{Greider01, Greider11}. 

A sporadic source of high-energy particles penetrating the Earth's atmosphere is related to solar eruptive processes, viz. solar flares, and coronal mass ejection (CMEs), where solar ions can be accelerated to high energies. Those particles are known as solar energetic particles (SEPs) \citep[e.g.][and references therein]{Cliver04, Desai2016}. The energy of SEPs is usually of the order of tens of MeV nucleon$^{-1}$, rarely exceeding 100 MeV nucleon$^{-1}$, but in some cases, SEPs can be accelerated to about GeV nucleon$^{-1}$ or even greater energy. In this case, similarly to the GCRs, SEPs produce a cascade of secondary particles in the Earth's atmosphere, that reaches the ground and increases the count rates of ground-based detectors, such as neutron monitors (NMs) \citep{Hat71, Greider01}. This special class of SEP events is called ground-level enhancements (GLEs) \citep[e.g.][]{Shea82, Poluianov2017}. The occurrence rate of GLEs is roughly ten per solar cycle, with a slight increase during the maximum and decline phase of the cycle \citep{Shea90, Stoker1995327, Klein20171107}.  

Accelerated to high energy solar ions lead to various space weather effects \citep[e.g.][]{Lilensten2009, Koskinen20171137}. SEPs lead to solar array performance degradation, harm on electronic components in space missions or single event effects leading to significant disruption of spacecraft performance. SEPs also pose a threat to astronauts as well as aircrews over transpolar flights \citep[e.g.][and references therein]{Vainio2009}. Therefore, SEPs, including GLE particles represent a specific and important space weather topic \citep[e.g.][and references therein]{Mishev2019swsc}. 

SEP and GCR fluxes, can be conveniently measured by space-borne instruments \citep[e.g.][]{Aguilar2010329, Adriani2016}. However, most of the space-borne instruments are constrained in the weight and size of the detector(s), which can affect their performance. Besides, space-borne probes are located most of the time in regions with high rigidity cut-off, which makes them poorly suitable for the study of SEPs. GLEs can be studied using the worldwide NM network \citep{Sim53, Hat71, Stoker2000, Mav11, Moraal201285, Papaioannou2014423}. 

Here, We propose an extension of the global neutron monitor network with several new detectors in order to optimize its performance, specifically for space weather purposes. We briefly discuss the ability of the current and optimized NM network for space weather services.  

\section{Plan for extension of the global NM network}
A NM is a complex ground-based detector aiming for registration of secondary particles, mostly neutrons, but also protons and a small amount of muons, produced by a primary CR particle in the Earth's atmosphere \citep{Sim57, Clem00}. Standard NM consists of sensitive to thermal neutrons proportional counters based on $^{3}$He or boron-trifluoride enriched to $^{10}$B, surrounded by a moderator, usually paraffin wax or polyethylene, a reflector made of the same material as the moderator and a lead producer \citep[for details see][and references therein]{Clem00, Simpson2000, Butikofer2018b}. The purpose of the moderator is to slow down, i.e., to reduce the energy of neutrons, leading to a considerable increase in their registration probability. The energy loss of a neutron during elastic collision increases with decreasing the atomic mass, therefore the moderator is selected to contain a significant amount of low mass nuclei e.g. Hydrogen. The lead producer, surrounds the moderator, aiming production of more neutrons by inelastic interactions in a thick target. Therefore, the producer is built by high atomic mass material. The outermost layer of the NM represents a moderator, namely the reflector, which has a double purpose: first, it rejects the low energy neutrons result from interaction(s) of the very local surroundings from penetrating in the NM, secondly, it allows to keep the produced in the lead neutrons inside the monitor.

The introduction of a NM as a continuous recorder of CR intensity followed the design by \citet{Sim53}. During the International Geophysical Year (IGY) 1957-1958 a 12 tube neutron monitor was constructed, but other configurations have been also used  \citep{Sim57, SheaSSR2000, Simpson2000}. The IGY neutron monitor was used world-wide as a detector to study CR variations. Lately, in the mid-sixties, the design of the IGY NM was optimized resulting in increased counting rate \citep{Hat64, Carmichael1968, Hat71}. This second generation of NM design is known as NM64 or supermonitor \citep[for details see][and references therein]{Simpson2000, Stoker2000}. Recently, mini-NMs have been installed at several stations, exhibiting good performance, specifically at low cut-off rigidity and high-altitude locations \citep{Poluianov2015281}.

The count rate of a NM provides reliable information about CR flux variations at the top of the Earth's atmosphere, both long-term (e.g. the 11-year sunspot cycle and the 22-year solar magnetic cycle), and short-term as Forbush decreases, diurnal CR variations and transient phenomena such as recently observed anisotropic cosmic ray enhancements \citep[for details see][]{Gil18}. NMs data are used to derive spectral and angular characteristics of GLEs and high-energy SEPs, specifically in the high-energy range and over the whole event timespan \citep[e.g.][]{Shea82, Cramp97, Bom06, Vas06, Mishev14c, Mishev2017swsc, Mishev2018sol72}. The information retrieved from NMs is essential to assess important topics related to space weather,  such as exposure to radiation of aircrew(s), henceforth exposure, and the influence of CRs on atmospheric chemistry \citep[e.g.][]{Bazilevskaya08, Vainio2009, Uso11b, Mironova2015}.

In order to offer a useful tool, specifically for space weather purposes, the global NM network shall provide coverage of the entire sky and real-time data access \citep[e.g.][]{Mav11}. Here, we discuss the current status of the global NM network and present a plan for its extension, aiming an optimization of its performance as a space weather tool.   
    
\subsection{Performance and current status of the global neutron monitor network}
Over the years, it was demonstrated that the global NM network is a powerful tool to study primary CR variations, transient phenomena, SEPs, and to provide data, which form an important input for space weather applications \citep[e.g.][]{Butikofer2018b}. In reality, the NM network as a whole, together with the geomagnetic field, represents a giant spectrometer, which allows one to observe the variations of the primary CRs, because NMs placed at various rigidity cut-offs are sensitive to different parts of CR spectrum. In addition, multi-vantage-point registration, specifically of SEPs, makes it possible to reveal the anisotropy of CRs in the vicinity of Earth, since the viewing cone of each station is a function on its location, particle rigidity, and angle of incidence of the arriving particle.

The global NM network presently consists of about 50 stations spread over the world, for details see Fig.1, where the NM stations with the corresponding rigidity cut-off are shown \citep{Moraal2000285, Mav11}. Here the computation of the rigidity cut-off over the globe was performed with the MAGNETOCOSMICS code using the IGRF magnetospheric model corresponding to the epoch 2015 \citep{Desorgher05, Thebault2015}.

The sensitivity of a NMs to primary CR is determined by the geomagnetic and atmospheric shielding. The rigidity cut-off is a function of the geomagnetic location of the monitor, while the thickness of the atmospheric layer above a given NM determines the atmospheric cut-off, since the primary CR must possess minimum energy ($\approx$430 MeV nucleon$^{-1}$ for the sea level) to induce an atmospheric cascade, whose secondary particles reach the ground \citep[e.g.][]{Greider01}. The atmospheric cut-off plays an important role in polar NMs, specifically those at the sea level, since the geomagnetic rigidity cut-off is small in the polar regions. Several high-altitude polar NMs, e.g. SOPO/SOPB and DOMC/DOMB are more sensitive to primary CR, specifically SEPs, than mid- and high rigidity cut-off NMs. Therefore, the rigidity range of the global NM network is determined by the atmospheric cut-off at polar regions, which posses the lower rigidity cut-offs, accordingly by the highest geomagnetic cut-off at about 17 GV in the magnetic Equator. 

Besides, polar NMs possess better angular resolution, which is important for the GLE analysis. With this in mind, a concept of the spaceship Earth, an optimized network consisting only of polar stations was proposed \citep{Bieber95}. However, one can see that the present NM network provides good coverage of arrival directions, and almost symmetric response (see Fig.1), but several gaps exist, as discussed below.

\begin{figure}[H]
   \centering
   \includegraphics[width=0.8\textwidth]{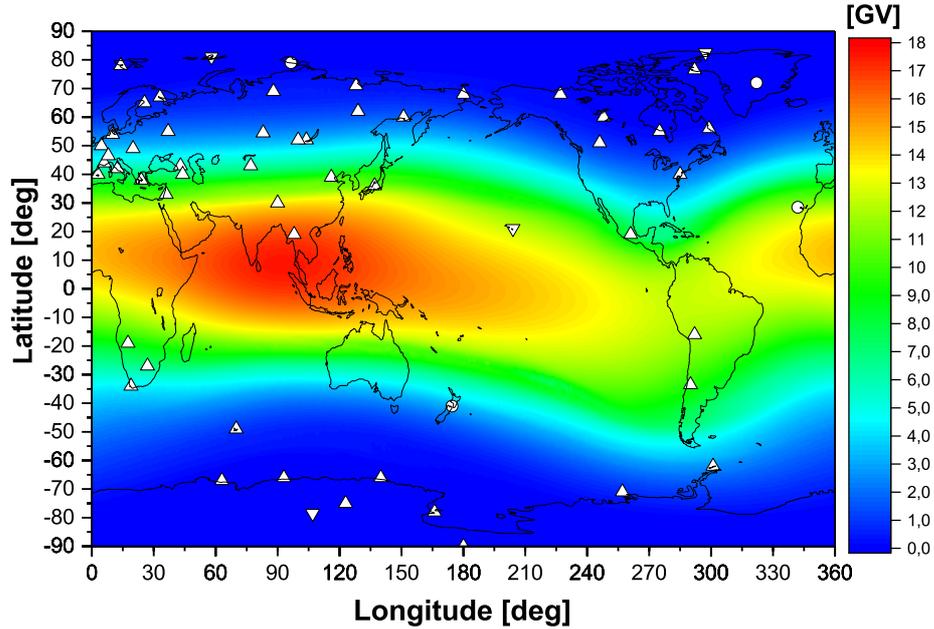}
      \caption{Present status of the global neutron monitor network and proposition for further extension. The up triangles correspond to presently operational stations. The down triangles correspond to previously existed stations. Circles correspond to the new stations proposed here. The color diagram depicts rigidity cut-off map computed in quiet magnetospheric conditions employing the IGRF model corresponding to epoch 2015 \citep{Thebault2015}.
              }
         \label{Fig1}
   \end{figure}

\subsection{Extension of the global NM network}
High-energy CRs are not deflected by the Earth's magnetic field. Therefore, NMs can record high-energy CRs, propagating almost along a straight line, determined by the latitude and longitude of the geographic position of the station. The situation is more complicated for low-energy particles, which are more strongly deflected. Thus, a NM is characterized by his asymptotic direction, i.e., the direction from which particles impinge on a given point in the atmosphere of the Earth arriving at the border of the magnetosphere. It depends on the location, particle incidence angle and rigidity \citep[for details see][and references therein]{Butikofer2018a, Butikofer2018b}. As a result, a NM is sensitive to a certain segment of the sky. While for the continuous recording of the isotropic GCR intensity, the asymptotic direction of a NM is not important, it is crucial for registration of GLEs, because SEPs reveal essential anisotropy, specifically during the event onset. Therefore, gaps in asymptotic directions of the global NM network can compromise the registration of GLEs, accordingly the corresponding analysis and alert services. 

The present situation of operational polar NMs  allows one to derive a comprehensive picture of GLE characteristics and provide alert systems (see Figs.1, 2 and Table 1). However, a gap in the asymptotic directions of Arctic NMs is observed, precisely in the longitude range 130--250$^{\circ}$ in the northern polar region. While the South polar NMs provide good coverage of the sky, those at North exhibit gaps (Fig.2). One can see that the majority of NMs are looking towards the Equator, i.e., NMs in the North hemisphere are looking southward, while those in Antarctica except DOMC, are looking northward. In addition, as was recently discussed, the high-altitude polar NMs such as DOMC and VSTK are more sensitive to SEPs \citep{Poluianov2017}. Therefore, there is a need for a NM, which is a counterpart of DOMC, i.e., high-altitude, low rigidity cut-off NM located in the North hemisphere close to the geomagnetic pole, as well as several stations to cover the gap and/or to improve the sensitivity, specifically in a low energy range. 

For example, if a GLE with narrow angular distribution of the particle flux occurs (see the pitch angle distribution in Fig.3) with anisotropy axis located in the polar region of the northern hemisphere, e.g. at 150$^{\circ}$ E, it would  not be registered by the existing NMs, because the rapidly diminishing from the apparent arrival direction particle flux (see the contours of equal pitch angle which also depict the particle flux intensity in the upper panel of Fig.2 and the pitch angle distribution (PAD) of  GLE $\#$5 in Fig.3). According to the current definition, a GLE event is registered when there are simultaneous statistically  significant enhancements of the count rates of at least two  differently located NMs including at least one station  near to sea level and a corresponding enhancement in the proton flux measured by space-borne instrument(s) \citep[for details see][]{Poluianov2017}. Therefore, the global NM network could not see a possible event, similar to GLE $\#$5, which poses major space weather thread (the strongest recorded GLE) occurring in the northern hemisphere (see the upper panel of Fig.2).

The existing gap can be filled, by an extension of the NM network with a NM at Severnaya Zemlya (SEVZ) (for details see the lower panel of Fig. 2 and Table 1) and by reopening of the presently non-operational, but previously existed NMs: Alert (ALRT) and Heis Island (HEIS). In addition, as a counterpart of DOMC, we propose a possible location of new NM on the Summit polar station in the Greenland plateau (Table 1), whose asymptotic direction is also given in Fig.2. Such an extended network of polar stations  would provide almost global coverage in the maximal NM response rigidity range of 1--5 GV and nearly to symmetric response of NMs from both hemispheres. Here, the computations were performed with the PLANETOCOSMICS code employing the IGRF magnetospheric model corresponding to the epoch 2015 \citep{Desorgher05, Thebault2015}. 

The extension of the global NM network involved several steps:

-- Determination of the gaps in the current network and possible locations for new stations (we selected only places with an existing facility providing power supply and data transfer);

-- Computation of the asymptotic directions of the new stations;

-- Comparison of performance between current and extended network;

-- Estimation of the necessary funds and drafting the corresponding proposal; 

\begin{figure}[H]
   \centering
   \includegraphics[width=0.8\textwidth]{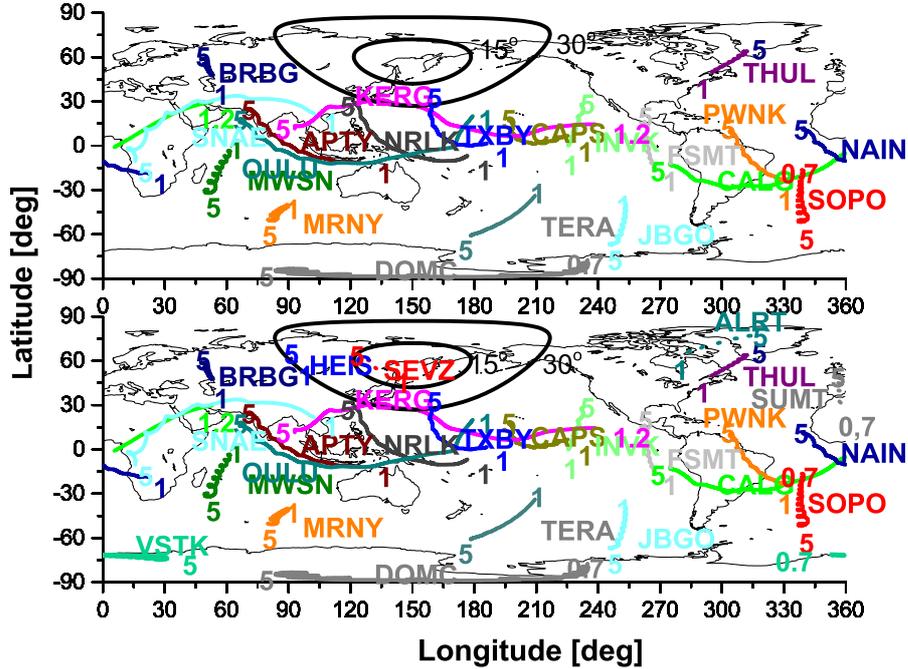}
      \caption{Asymptotic directions of polar NMs. The abbreviations are given in Table 1.
      The color lines depict asymptotic directions plotted in the rigidity range 1--5 GV, for DOMC, SOPO, SUMT and VSTK from 0.7 to 5 GV respectively. The dashed lines correspond to new NMs proposed for extension of the network or to be reopened. The lines of equal pitch angles relative to the anisotropy axis of example event are plotted for 15$^{\circ}$ and 30$^{\circ}$ for sunward direction. The upper panel corresponds to the current global NM network, while the lower panel to the extended NM network. The figure is adapted from \citep{Mishev18swsc}.
              }
         \label{Fig7}
   \end{figure}

\begin{table*}[htp]
\caption{Neutron monitors used in this study. Columns represent station name, location, geomagnetic cut-off rigidity and altitude above sea level. The table encompasses the current status of low rigidity stations (the part above the dashed line), the closed but previously existing stations to be reopened (the part between the dashed and dashed-dashed lines)  and new stations proposed to extend the network (the bottom part). }
\label{table:3}
\centering
\small

\begin{tabular}{c c c c c }     % 5 columns
%\hline\hline
                      % To combine 4 columns into a single one

Station & latitude [deg] & Longitude [deg] &  $P_{c}$ [GV] & Altitude [m] \\
\hline

   Apatity (APTY)& 67.55  & 33.33  & 0.57 &  177  \\
   Barenstburg (BRBG) & 78.03  & 14.13  & 0.01 & 51  \\
   Calgary (CALG) & 51.08 & 245.87  & 1.08  &  1128 \\
   Cape Schmidt (CAPS) & 68.92 & 180.53  & 0.45 &  0  \\
   Dome C (DOMC) & -75.06 & 123.20 & 0.01 & 3233   \\
   Forth Smith (FSMT)  & 60.02 & 248.07  & 0.381 &  0  \\
   Inuvik (INVK)     & 68.35  & 226.28  & 0.16  &  21 \\
   Jang Bogo(JNBG) & -74.37  & 164.13  & 0.1 & 29        \\
   Kerguelen (KERG) & -49.35 & 70.25 & 1.01 & 33          \\
   Mawson (MWSN) & -67.6 & 62.88 & 0.22 & 0               \\
   Mirny  (MRNY) & -66.55 & 93.02 & 0.03 & 30               \\
   Nain (NAIN) & 56.55 & 298.32 & 0.28 & 0                      \\
   Neumayer (NEUM) & -70.40 & 351.04 & 0.85 & 0                      \\
   Norilsk (NRLK) & 69.26 & 88.05 & 0.52 & 0  \\
   Oulu (OULU) & 65.05 & 25.47 & 0.69 & 15                       \\
   Peawanuck (PWNK) & 54.98 & 274.56 & 0.16 & 52     \\
   Sanae (SNAE) & -71.67 & 357.15 & 0.56 & 52  \\
   South Pole (SOPO) & -90.00 & 0.0 & 0.09 & 2820             \\
   Terre Adelie (TERA) & -66.67 & 140.02 & 0.02 & 45            \\
   Thule (THUL) & 76.60 & 291.2 & 0.1 & 260                      \\
   Tixie (TXBY) & 71.60 & 128.90 & 0.53 & 0                \\  
%   \hdashline
    Alert (ALRT)        & 82.5  & 297.67  & 0.0   & 57 \\
    Heiss island (HEIS) & 80.62 & 58.05   & 0.1   & 20  \\
    Haleakala (HLEA)   & 20.71  & 203.74  & 12.91 & 3052      \\ 
    Vostok (VSTK)      & -78.47 & 106.87  & 0.0   & 3488        \\   
%    \hdashline\hdashline
    Canary Islands (CANI)  & 28.45 & 342.47  & 11.76   & 2376        \\       
    New Zealand    (NZLD)  & -43.59 & 170.27 & 3.28    & 1029        \\       
    Severnaya Zemlya (SEVZ) & 79.29 & 96.5 & 0.11    & 10       \\       
    Summit (SUMT)          & 72.34 & 321.73 & 0.01    & 3126       \\       
\hline
\end{tabular}
\end{table*}

\section{Services and applications provided by the extended global NM network}
Here, We present several abilities of the global NM network, related to space weather services and solar physics research.

\subsection{Registration and analysis of GLEs}
Registration of a GLE can provide an early alert for the onset of SEP event, which is specifically important for various space weather services \citep[for details see][]{Kuwabara2006b, Kuwabara2006a}. Accordingly, alert systems, based on NM records have been developed \citep{Souvatzoglou2014633, Mavromichalaki20181797, Dorman20192490}. Most of those alert systems are based on a good coverage of the arrival direction of GLE particles by the global NM network since a given number of stations shall exhibit a count rate increase. Therefore, an extended global NM network will provide a reliable basis for the corresponding alert service(s). Besides, the spectral and angular characteristics of strong SEP events, viz. GLEs in the energy range $\sim$ 0.3--20 GeV nucleon$^{-1}$, can be derived by modeling of the global NM network response.

Methods for analysis of GLEs using NM data have been developed over the years, usually based on modeling of the global NM network response and optimization of a set of unknown model parameters $n$ over the experimental data points corresponding to the number of NM stations \citep[e.g.][]{Shea82, Cramp97, Bom06, Vas06}. In general, the relative count rate increase of a given NM during GLE can be expressed as:

\begin{equation}
\frac{\Delta N(P_{cut})}{N(t)} =\frac{\sum_{i}\sum_{k} \int_{P_{cut}}^{P_{max}}J_{sep_{i}}(P,t)S_{i,k}(P)G_{i}(\alpha(P,t)) A_{i}(P)dP}{\sum_{i}\int_{P_{cut}}^{\infty}J_{GCR_{i}}(P,t)S_{i}(P)dP}
\label{simp_eqn1}
   \end{equation}
\noindent where $N$ is the count rate due to GCR averaged over two hours before the event's onset \citep[e.g.][]{Usoskin2015}, which can be also variable in case of a long event occurred during a Forbush decrease, $\Delta N(P_{cut})$ is the count rate increase due to solar particles. $J_{sep}$ is the rigidity spectrum of $i$ (proton or $\alpha$-particle) component of SEPs, usually only protons are taken into account, accordingly $J _{GCR_{i}}(P,t)$ is the rigidity spectrum of the $i$ component (proton or $\alpha$-particle, etc...) of GCR at given time $t$, $G(\alpha(P,t))$ is the pitch angle distribution of SEPs, otherwise, for GCRs the angular distribution is assumed to be isotropic, accordingly, A(P) is a discrete function with $A(P)$=1 for allowed trajectories (proton with rigidity $P$ can reach the station) and $A(P)$=0 for forbidden trajectories (proton with rigidity $P$ cannot reach the station). Function $A$ is derived during the asymptotic cone computations. $P_{cut}$ is the minimum rigidity cut-off of the station, accordingly, $P_{max}$ is the maximum rigidity of SEPs considered in the model, whilst for GCR $P_{max}$= $\infty$. $S_{k}$ is the NM yield function for vertical and for oblique incidence SEPs \citep{Clem97}. The contribution of oblique SEPs to NM response is particularly important for modeling strong and/or very anisotropic events, while for weak and/or moderately strong events it is possible to consider only vertical ones and using $S_{k}$ for an isotropic case, which considerably simplifies the computations \citep{Mishev16SF}. 

The background due to GCRs can be computed using a convenient model, e.g., the force-field model with the corresponding local interstellar spectrum, considering explicitly the modulation potential \citep{Usoskin2005, Vos2015}. The optimization can be performed over the set of model parameters $n$ by minimizing the difference between the modeled and measured NM responses using a convenient method  \citep{Tikhonov1995, Mavrodiev2004359, Aster2005, Mishev20057016}.  The modeling of the global network NM response can be performed using the corresponding NM yield function, which establishes a connection between the primary CR flux at the top of the Earth's atmosphere and the count rate of the device. Since the secondary CRs, resulting from the primary CR induced cascade in the Earth's atmosphere, can reach the ground level and eventually be registered by a NM, the yield function incorporates the full complexity of the atmospheric cascade development including secondary particle propagation in the atmosphere and the efficiency of the detector itself to register the secondaries \citep[e.g.][and references therein]{Clem00}. The NM yield function can be determined by parameterization of experimental data, namely latitude survey(s) \citep[e.g.][]{Nagashima1989, Raubenheimer81, Dorman2000} or can be assessed using Monte Carlo simulations of CR propagation in the atmosphere  \citep[e.g.][]{Debrunner68, Clem00}. Recently, essential progress of Monte Carlo simulations of CR propagation in the atmosphere was achieved, which resulted in several newly computed NM functions \citep{Clem00, Flu08, Mishev13b, Mangeard20167435}. A recently computed NM yield function by \citet{Mishev13b, MishevNMYFJGR2020} is fully consistent with the experimental latitude surveys  and was validated by achieving good agreement between model results and measurements, including space-borne data  \citep{Gil15, Nuntiyakul20187181, koldobsky19}. 

As an example, we present the derived spectra and PAD of GLE $\#$5, which was the largest event ever observed by the global NM network. It occurred on  23 February 1956 and was registered by various ground-based detectors (ionization chambers, NMs and muon telescopes) and recently was reassessed  \citep[e.g.][]{Vashenyuk2008926, MishevJGR2020_GLE5}. This event was very anisotropic. Significant asymmetry between the count rate increases recorded by several European NMs, namely Leeds (LEED), Stockholm (STHM) and Weissenau (WEIS) and American ones, namely Chicago (CHGO), Calgary (CALG) and Ottawa (OTWA) was observed. The stations in Europe revealed rapid and very large NM count rate increases, while those in North America were with considerably delayed maximum and smaller count rate enhancements, for details see \texttt{gle.oulu.fi}. The derived SEPs spectra and PAD are shown in Fig.3. The relativistic solar proton spectra were very hard, specifically during the event's onset initial phase, whilst a narrow PAD was revealed. The SEP spectra remained hard (with nearly exponential shape) during the whole event, in contrast to other GLEs \citep[e.g.][and references therein]{Miroshnichenko18}.

The extended NM network allows to significantly improve the optimization procedure, namely it results in reduction of the residual $\mathcal{D}$, which is defined as:

\begin{equation}
\mathcal{D}=\frac{\sqrt{\sum_{i=1}^{m} \left[\left(\frac{\Delta N_{i}}{N_{i}}\right)_{mod.}-\left(\frac{\Delta N_{i}}{N_{i}}\right)_{meas.}\right]^{2}}}{\sum_{i=1}^{m} (\frac{\Delta N_{i}}{N_{i}})_{meas.}}
\label{simp_eqn7}
   \end{equation} 
   
\noindent where $m$ is the number of NM stations, $\frac{\Delta N_{i}}{N_{i}}$ is the relative NM count rate increase for the $i$ NM station. 

A robust optimization process and reliable solution are achieved when $\mathcal{D}$ $\le$ 5 $\%$, a criterion usually fulfilled for strong events, whilst for moderately strong and weak events $\mathcal{D}$ can be about 8--12 $\%$. We emphasize that a solution can be obtained even in the case of $\mathcal{D}$ $\sim$ 20--30 $\%$, but with considerably larger uncertainties. Usually, it is necessary to possess about 2(n-1) data (NM stations), $n$ is the number of unknowns in the model, in order to be able to unfold the model parameters \citep[e.g.][]{Himmelblau72, Den83, Mavrodiev2004359}. Thus, it is sufficient to retrieve information from 15--20 NMs, specifically those in a polar region, whilst the mid-latitude stations provide the boundary conditions. However, this number of stations is reasonable in case of not complicated PAD and unidirectional SEP flux, such as  GLE $\#$ 59 or  GLE $\#$ 70 \citep[for details see][]{Mishev16SF, Mishev2017swsc}. In case of more complicated PADs and/or bi-directional SEP flux, e.g., GLE $\#$69 or GLE $\#$71 \citep[for details see][]{Mishev14c, Mishev2018sol72}, the amount of required information considerably increases, leading to about 30--35 NM records necessary to perform a reliable analysis.  

Here, we examined the performance of the extended, actual and reduced NM network for an analysis of several GLEs, the details are given in Table 2. One can see that the extended NM network results in a notably smaller $\mathcal{D}$ compared to the actual number of NMs used for the analysis, whilst a reduction of the number of NMs leads to a considerable reduction of the ability of the global NM network to provide a reliable GLE analysis. The additional data used for the analysis with the extended NM network are based on forward modeling including realistic noise similarly to \citet{Mavrodiev2004359} employing the derived spectra and PAD during the actual analysis. We note, that the extended analysis is performed with all polar stations from Table 1, which encompasses the extended network,  who are added to the actual analysis (a partial overlapping exists for some events, since  several NMs from Table 1 are used also for the actual analysis). For the analysis with the reduced NM network we removed about 5--10 NMs with moderate response.            

\begin{figure}[H]
   \centering
   \includegraphics[width=0.8\textwidth]{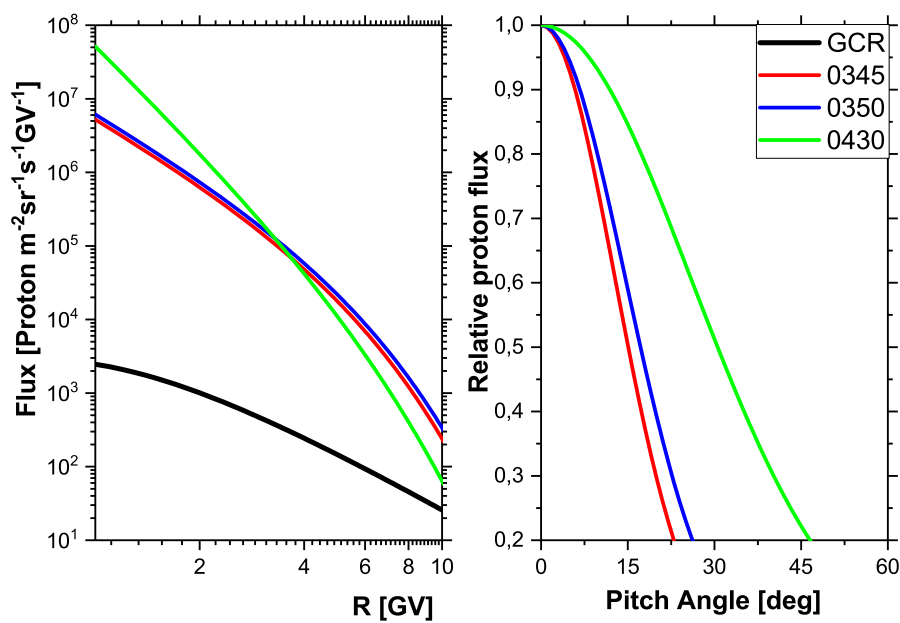}
      \caption{Derived rigidity spectra and PAD during GLE $\#$5 at various stages of the events specified in UT time in the legend. Left panel corresponds to the derived rigidity spectra, while the right panel corresponds to the corresponding pitch angle distribution.
              }
         \label{Fig3}
   \end{figure}

\begin{table*}[htp]
\caption{The value of the merit function $\mathcal{D}$ obtained for the analysis of several GLEs (main phase of the event) as a function of the number of the used NM stations. Columns 1--2 correspond to the number and date of the GLE, while columns 3--5 correspond to $\mathcal{D}$ and number of the used stations (in the brackets) for extended NM network, actual NM network used for the analysis and the reduced NM network, respectively. N.A. depicts the case when the SEP spectra  cannot be unfolded. The details for the analysis of the presented  GLEs are given in \citep{Mishev14c, Mishev16SF, Kocharov2017, Mishev2018sol72} as well as in this work.}
\label{table:4}
\centering
\small

\begin{tabular}{c c| c c c}     % 5 columns
%\hline\hline
                      % To combine 4 columns into a single one

GLE $\#$     &  Date   &  Extended NM network  & Actual NM network  & Reduced nNM etwork \\
\hline
   GLE $\#$ 5  & 23.02.1956  & 1.6(37)  & 2.5(15)    & N.A.(10)  \\  
   GLE $\#$ 59 & 14.07.2000  & 4.1(39)  & 4.8(30)    & 19(20)     \\
   GLE $\#$ 67 & 02.11.2003  & 4.5(39)  & 7.1(34)    & 38(21)     \\
   GLE $\#$ 69 & 20.01.2005  & 3.0(38)  & 3.5(33)    & 35(25)     \\ 
   GLE $\#$ 70 & 13.12.2006  & 3.2(38)  & 4.2(32)    & 43(22)     \\
   GLE $\#$ 71 & 17.05.2012  & 5.0(34)  & 7.1(24)    & N.A.(19)   \\   
   GLE $\#$ 72 & 10.09.2017  & 5.2(31)  & 6.1(23)    & 33(18)     \\

\end{tabular}
\end{table*}   
   
\subsection{Space weather purposes - exposure during GLEs}
The increased intensity of CRs during SEP events, leads to an important space  weather issue, namely exposure at flight altitudes \citep[e.g.][and references therein]{Mewaldt2006303, Pulkkinen20071, Shea12}. During intercontinental flights over the sub-polar and polar regions, aircrews are exposed to non-negligible radiation field due to secondary particles, which can be significantly enhanced during major GLEs \citep{Spurny96, Spurny02, Shea2000}. Assessments of the exposure during GLEs requires detailed information of SEP spectra as an input for a relevant model for computation of the exposure \citep[e.g.][]{Ferrari01, Latocha2009286, Copeland2017419}.  

Here we present as an example the exposure to radiation at flight altitude during the strongest ever observed GLE. The computation was  performed using a numerical model \citep{Mishev2015, Mishev18swsc}. The effective dose rate at a given atmospheric depth $h$ induced by a primary CR particle is computed by convolution of the exposure yield function with the corresponding primary CR particle spectrum:
\begin{equation}\label{eq:1}
E(h, T, \theta, \varphi) = \sum_{i}\int_{T(P_{cut})}^{\infty}\int_{\Omega}J_i(T)Y_i(T, h)d\Omega(\theta, \varphi) dT,
\label{simp_eqn5}
\end{equation}
where $J_{i}(T)$ is the differential energy spectrum of the primary CR arriving at the top of the atmosphere for $i-$th component
 (proton or $\alpha-$particle) and $Y_{i}$ is the effective dose yield function for this type of particles.
The integration is over the kinetic energy $T$ above $T_{cut}(P_{c})$, which is defined by the local cut-off rigidity $P_{c}$
 for a nucleus of type $i$, $T_{cut,i}=\sqrt{ \left( \frac {Z_{i}} {A_{i}}\right)^{2} P_{c}^{2}+ E_{0}^{2}} - E_{0}$,
 where $E_{0}$ =  0.938 GeV/c$^{2}$ is the proton's rest mass.

Accordingly, the effective dose  yield function  $Y_{i}$ is:
\begin{equation}
Y_{i}(T,h) = \sum_{j} \int\limits_{T^{*}}  F_{i,j}(h,T,T^{*}, \theta,\varphi) C_{j}(T^{*}) dT^{*}
\label{simp_eqn6}
   \end{equation}
where $C_{j}(T^{*})$ is the coefficient converting the fluence of secondary particles of type $j$ (neutron, proton, $\gamma$, $e^{-}$,
 $e^{+}$, $\mu^{-}$, $\mu^{+}$, $\pi^{-}$, $\pi^{+}$) with energy $T^{*}$ to the effective dose, $F_{i,j}(h,T,T^{*}, \theta,\varphi)$
 is the fluence of secondary particles of type $j$, produced by a primary particle of type $i$ (proton or $\alpha-$particle)
 with given primary energy $T$ arriving at the top of the atmosphere from zenith angle $\theta$ and azimuth angle $\varphi$. 
The conversion coefficients $C_{j}(T^{*})$ are considered according to \citet{Petoussi2010}. We note, that employment of different conversion coefficients $C_{j}(T^{*})$ \citep[e.g.][]{ICRP1996}, would lead to increase of the exposure assessment of about 20 $\%$, which is considerably below the other model uncertainties \citep[e.g.][]{Copeland2019GLE, Yang2020}. 

Using the derived rigidity spectra for GLE $\#$5 (Fig.3) and Eq. (3), we computed the effective dose rate at a typical altitude for an intercontinental commercial jet flight of 40 kft (12 190m) a.s.l., altitude representative for a polar flight over a polar atmosphere \citep[for details see][and references therein]{Mishev2010476, Mishev14d, Mironova2015}, similarly to  \citet{Mishev20181921, Copeland2019GLE}. Here, we would like to stress that the exposure during GLEs can usually reach peak values considerably greater than the GCR background, but for a relatively short period. Therefore, it is more relevant to integrate the exposure over a certain period, naturally related to the flight duration. However, during GLE $\#$5, the derived SEP spectra remained hard even after the event initial and main phase of the event, i.e., for a relatively long period, which is comparable with a polar flight duration. The distribution of the effective dose over the globe at an altitude of 40 kft a.s.l., integrated over the first four hours after the event onset during GLE $\#$5 is presented in Fig.4. One can see that the exposure is significant in a polar region, where the received dose is considerably greater than the suggested annual limit for occupational workers of about 6 mSv \citep[e.g.][]{Euratom13}. The received dose for the population integrated over four hours in the polar region, which is a typical time span of flight in this region, is about an order of magnitude greater than the recommended of 1 mSv \citep[e.g.][]{Euratom13}. The accumulated exposure is significant even at mid- and high-rigidity cut-off regions, because of the very hard SEP spectra. 

\begin{figure}[H]
   \centering
   \includegraphics[width=0.8\textwidth]{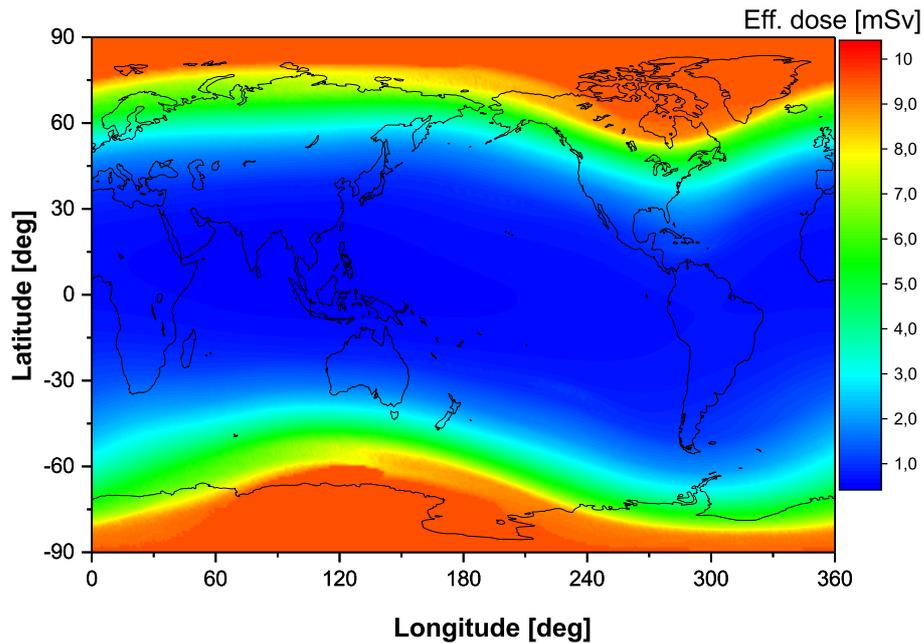}
      \caption{Global map of the  effective dose at altitude of 40 kft during GLE $\#$5 integrated over the first four hours of the event.
              }
         \label{Fig4}
   \end{figure}   

\subsection{Registration of solar neutrons}
The global NM network provides a good opportunity to study solar neutrons \citep[e.g.,][]{Usoskin1997, Dorman2010, Artamonov2016}. During solar eruptions, accelerated high-energy ions can interact with matter in the solar atmosphere, resulting in in-situ production of different types of secondary particles, e.g. $\gamma$-rays and neutrons \citep[for details see][and references therein]{Hurford2003, Dorman2010}. Of specific interest are neutrons, the so-called solar neutrons \citep[e.g.,][and references therein]{Lingenfelter1965a}. Since the solar neutrons are neutral, they propagate straight to the Earth, therefore bringing direct information of the acceleration site. If the energy of solar neutrons is greater than about 100 MeV, they can induce a nucleonic cascade in the Earth's atmosphere and can be registered by NMs. The sensitivity of a NM to solar neutrons is greater when the atmospheric depth in the solar direction is smaller, because the atmosphere attenuates the flux of secondary nucleons in the cascade. An optimal location is high-altitude, close to the Equator \citep{Usoskin1997}. In order to improve this capability, it is recommended to extend the current network with at least two high-altitude NMs, namely one located at the Canary Islands and the other in New Zealand, and to re-open the Haleakala (HLEA) NM, details given are in Table 1, \citep[see also][]{Artamonov2016}. We note that the Canary Island NM is under construction (Private communication).

\section{Conclusions}
We discussed the current status and application of the global neutron monitor network to study solar energetic particles, specifically for space weather purposes, namely alerts, assessment of SEP characteristics and the corresponding computation of the exposure to radiation at flight altitudes. 

As an example,  we presented the ability of the global NM network data to be used for derivation of the spectra and angular distribution of SEPs during the strongest GLE event of the observational era: GLE $\#$5 and the related in the course of the event effective dose over the globe. In order to improve those capabilities, we propose  to reopen four previously operational NMs, namely ALRT, HEIS, HLEA and VSTK (see Table 1, stations below the dashed line) and to build four new stations: CANI, NZLD, SEVZ, SUMT (see Table 1, stations below the dashed-dashed line). Hence, covering several existing gaps and improving its sensitivity specifically in the low energy range, the global NM network will be a useful tool to study various populations of solar particles and will be a useful instrument for space weather services. 

Besides, in order to keep operational those capabilities of the global NM network, we would like to stress that even a partial reduction of the number of existing NMs would considerably influence the usage of the global NM network as a convenient tool for space weather services. Since at present the existence and continuous functioning of several NM stations are under question, the support of the network from governments, foundations(s) and space flight operators is crucially needed. 

%%%%%%%%%%%%%%%%%%%%%%%%%%%%%%%%%%%%%%%%%%%%%%%%%%%%%%%%%%%%%%%%%%%%%%%%%%%
\section*{Acknowledgements}
This work was supported by the Academy of Finland (project 321882 ESPERA) and (project 304435 CRIPA-X). The work benefits from discussions in the framework of the International Space Science Institute International Team 441: High EneRgy sOlar partICle Events Analysis (HEROIC). The authors acknowledge all the researchers, NM station managers and colleagues who collected the GLE records used for the analysis of GLE $\#$5: ALBQ, ARNB, BERK, CHGO, CLMX, GOTT, HUAN, LEED, MTNR, MTWL, MXCO, OTWA, SACR, STHM, WEIS. The NM data were retrieved from the international GLE database (\texttt{http://gle.oulu.fi/$\#$/}). Oulu NM data are also available at \texttt{http://cosmicrays.oulu.fi}.

%\acknowledgment US spelling: \verb+\acknowledgment+
%\acknowledgement British  spelling: \verb+\acknowledgement+

%%%%%%%%%%%%%%%%%%%%%%%%%%%%%%%%%%%%%%%%%%%%%%%%%%%%%%%%%%%%%%%%%%%%%%%%%%%

%\end{article} 

\end{document}